\batchmode
\documentstyle[12pt]{article}
\newcommand{\nc}{\newcommand}
\nc{\renc}{\renewcommand}

\nc{\half}{{\textstyle{1\over2}}}
\nc{\etal}{\mbox{\it et al. }}
\nc{\ie}{{\it i.e.}}
\nc{\eg}{{\it e.g.}}

\renc{\thefootnote}{\arabic{footnote}}
\nc{\capt}[1]{{\bf Figure.} {\small\sl #1}}

\nc{\eqs}[2]{\mbox{Eqs.~(\ref{#1},\,\ref{#2})}}
\nc{\eq}[1]{\mbox{Eq.~(\ref{#1})}}

\nc{\figs}[2]{\mbox{Figs.~(\ref{#1},\,\ref{#2})}}
\nc{\fig}[1]{\mbox{Fig~.(\ref{#1})}}

\nc{\tag}[1]{\label{#1} \marginpar{{\footnotesize #1}}}
\nc{\mtag}[1]{\label{#1} \mbox{\marginpar{{\footnotesize #1}}}}
\renc{\baselinestretch}{1.5}
\newlength{\overeqskip}
\newlength{\undereqskip}
\nc{\be}[1]{\begin{equation} \mbox{$\label{#1}$}}
\nc{\bea}[1]{\begin{eqnarray} \mbox{$\label{#1}$}}
\nc{\Section}[2]{\section{#2}\label{#1}}
\nc{\Bibitem}[1]{\bibitem{#1}}
\nc{\Label}[1]{\label{#1}}

\nc{\eea}{\vspace{\undereqskip}\end{eqnarray}}
\nc{\ee}{\vspace{\undereqskip}\end{equation}}
\nc{\bdm}{\begin{displaymath}}
\nc{\edm}{\end{displaymath}}
\nc{\dpsty}{\displaystyle}
\nc{\bc}{\begin{center}}
\nc{\ec}{\end{center}}
\nc{\ba}{\begin{array}}
\nc{\ea}{\end{array}}
\nc{\bab}{\begin{abstract}}
\nc{\eab}{\end{abstract}}
\nc{\btab}{\begin{tabular}}
\nc{\etab}{\end{tabular}}
\nc{\bit}{\begin{itemize}}
\nc{\eit}{\end{itemize}}
\nc{\ben}{\begin{enumerate}}
\nc{\een}{\end{enumerate}}
\nc{\bfig}{\begin{figure}}
\nc{\efig}{\end{figure}}
\nc{\arreq}{&\!=\!&}
\nc{\arrmi}{&\!-\!&}
\nc{\arrpl}{&\!+\!&}
\nc{\arrap}{&\!\!\!\approx\!\!\!&}
\nc{\non}{\nonumber\\*}
\nc{\align}{\!\!\!\!\!\!\!\!&&}

\def\lsim{\; \raise0.3ex\hbox{$<$\kern-0.75em
      \raise-1.1ex\hbox{$\sim$}}\; }
\def\gsim{\; \raise0.3ex\hbox{$>$\kern-0.75em
      \raise-1.1ex\hbox{$\sim$}}\; }
\nc{\DOT}{\hspace{-0.08in}{\bf .}\hspace{0.1in}}
\nc{\Laada}{\hbox {$\sqcap$ \kern -1em $\sqcup$}}
\nc\loota{{\scriptstyle\sqcap\kern-0.55em\hbox{$\scriptstyle\sqcup$}}}
\nc\Loota{{\sqcap\kern-0.65em\hbox{$\sqcup$}}}
\nc\laada{\Loota}
\nc{\qed}{\hskip 3em \hbox{\BOX} \vskip 2ex}

\nc{\real}{{\rm I \! R}}
\nc{\Z}{{\sf Z \!\!\! Z}}
\nc{\complex}{{\rm C\!\!\! {\sf I}\,\,}}
\def\bigid{\leavevmode\hbox{\small1\kern-3.8pt\normalsize1}}
\def\id{\leavevmode\hbox{\small1\kern-3.3pt\normalsize1}}
\nc{\slask}{\!\!\!/}
\nc{\bis}{{\prime\prime}}
\nc{\pa}{\partial}
\nc{\na}{\nabla}
\nc{\ra}{\rangle}
\nc{\la}{\langle}
\nc{\goto}{\rightarrow}
\nc{\swap}{\leftrightarrow}

\nc{\EE}[1]{ \mbox{$\cdot10^{#1}$} }
\nc{\abs}[1]{\left|#1\right|}
\nc{\at}[2]{\left.#1\right|_{#2}}
\nc{\norm}[1]{\|#1\|}
\nc{\abscut}[2]{\Abs{#1}_{\scriptscriptstyle#2}}
\nc{\vek}[1]{{\rm\bf #1}}
\nc{\integral}[2]{\int\limits_{#1}^{#2}}
\nc{\inv}[1]{\frac{1}{#1}}
\nc{\dd}[2]{{{\partial #1}\over{\partial #2}}}
\nc{\ddd}[2]{{{{\partial}^2 #1}\over{\partial {#2}^2}}}
\nc{\dddd}[3]{{{{\partial}^2 #1}\over
        {\partial #2 \partial #3}}}
\nc{\dder}[2]{{{d #1}\over{d #2}}}
\nc{\ddder}[2]{{{d^2 #1}\over{d {#2}^2}}}
\nc{\dddder}[3]{{d^2 #1}\over
        {d #2 d #3}}
\nc{\dx}[1]{d\,^{#1}x}
\nc{\dy}[1]{d\,^{#1}y}
\nc{\dz}[1]{d\,^{#1}z}
\nc{\dl}[1]{\frac{d\,^{#1}l}{(2\pi)^{#1}}}
\nc{\dk}[1]{\frac{d\,^{#1}k}{(2\pi)^{#1}}}
\nc{\dq}[1]{\frac{d\,^{#1}q}{(2\pi)^{#1}}}

\nc{\cc}{\mbox{$c.c.$ }}
\nc{\hc}{\mbox{$h.c.$ }}
\nc{\cf}{cf.\ }
\nc{\erfc}{{\rm erfc}}
\nc{\Tr}{{\rm Tr\,}}
\nc{\tr}{{\rm tr\,}}
\nc{\pol}{{\rm pol}}
\nc{\sign}{{\rm sign}}
\nc{\bfT}{{\bf T }}
\def\eV{{\rm\ eV}}
\def\GeV{{\rm\ GeV}}
\def\MeV{{\rm\ MeV}}

\def\TeV{{\rm\ TeV}}

\nc{\cA}{{\cal A}}
\nc{\cB}{{\cal B}}
\nc{\cD}{{\cal D}}
\nc{\cE}{{\cal E}}
\nc{\cG}{{\cal G}}
\nc{\cH}{{\cal H}}
\nc{\cL}{{\cal L}}
\nc{\cO}{{\cal O}}
\nc{\cT}{{\cal T}}
\nc{\cN}{{\cal N}}
\nc{\rvac}[1]{|{\cal O}#1\rangle}
\nc{\lvac}[1]{\langle{\cal O}#1|}
\nc{\rvacb}[1]{|{\cal O}_\beta #1\rangle}
\nc{\lvacb}[1]{\langle{\cal O}_\beta #1 |}
\nc{\bb}{\bar{\beta}}
\nc{\bt}{\tilde{\beta}}
\nc{\ctH}{\tilde{\cal H}}
\nc{\chH}{\hat{\cal H}}
\nc{\1}{\aa}
\nc{\2}{\"{a}}
\nc{\3}{\"{o}}
\nc{\4}{\AA}
\nc{\5}{\"{A}}
\nc{\6}{\"{O}}
\nc{\al}{\alpha}
\nc{\g}{\gamma}
\nc{\Del}{\Delta}
\nc{\e}{\epsilon}
\nc{\eps}{\epsilon}
\nc{\lam}{\lambda}
\nc{\om}{\omega}
\nc{\Om}{\Omega}
\nc{\ve}{\varepsilon}
\nc{\mn}{{\mu\nu}}
\nc{\k}{\kappa}
\nc{\vp}{\varphi}

\nc{\advp}[3]{{\it  Adv.\ in\ Phys.\ }{{\bf #1} {(#2)} {#3}}}
\nc{\annp}[3]{{\it  Ann.\ Phys.\ (N.Y.)\ }{{\bf #1} {(#2)} {#3}}}
\nc{\apl}[3]{{\it  Appl. Phys. Lett. }{{\bf #1} {(#2)} {#3}}}
\nc{\apj}[3]{{\it  Ap.\ J.\ }{{\bf #1} {(#2)} {#3}}}
\nc{\apjl}[3]{{\it  Ap.\ J.\ Lett.\ }{{\bf #1} {(#2)} {#3}}}
\nc{\app}[3]{{\it Astropart.\ Phys.\ }{{\bf #1} {(#2)} {#3}}}
\nc{\cmp}[3]{{\it  Comm.\ Math.\ Phys.\ }{{ \bf #1} {(#2)} {#3}}}
\nc{\cqg}[3]{{\it  Class.\ Quant.\ Grav.\ }{{\bf #1} {(#2)} {#3}}}
\nc{\epl}[3]{{\it  Europhys.\ Lett.\ }{{\bf #1} {(#2)} {#3}}}
\nc{\ijmp}[3]{{\it Int.\ J.\ Mod.\ Phys.\ }{{\bf #1} {(#2)} {#3}}}
\nc{\ijtp}[3]{{\it Int.\ J.\ Theor.\ Phys.\ }{{\bf #1} {(#2)} {#3}}}
\nc{\jmp}[3]{{\it  J.\ Math.\ Phys.\ }{{ \bf #1} {(#2)} {#3}}}
\nc{\jpa}[3]{{\it  J.\ Phys.\ A\ }{{\bf #1} {(#2)} {#3}}}
\nc{\jpc}[3]{{\it  J.\ Phys.\ C\ }{{\bf #1} {(#2)} {#3}}}
\nc{\jap}[3]{{\it J.\ Appl.\ Phys.\ }{{\bf #1} {(#2)} {#3}}}
\nc{\jpsj}[3]{{\it J.\ Phys.\ Soc.\ Japan\ }{{\bf #1} {(#2)} {#3}}}
\nc{\lmp}[3]{{\it Lett.\ Math.\ Phys.\ }{{\bf #1} {(#2)} {#3}}}
\nc{\mpl}[3]{{\it  Mod.\ Phys.\ Lett.\ }{{\bf #1} {(#2)} {#3}}}
\nc{\ncim}[3]{{\it  Nuov.\ Cim.\ }{{\bf #1} {(#2)} {#3}}}
\nc{\np}[3]{{\it  Nucl.\ Phys.\ }{{\bf #1} {(#2)} {#3}}}
\nc{\npps}[3]{{\it  Nucl.\ Phys.\ Proc.\ Suppl.\ }{{\bf #1} {(#2)} {#3}}}
\nc{\pr}[3]{{\it Phys.\ Rev.\ }{{\bf #1} {(#2)} {#3}}}
\nc{\pra}[3]{{\it  Phys.\ Rev.\ A\ }{{\bf #1} {(#2)} {#3}}}
\nc{\prb}[3]{{\it  Phys.\ Rev.\ B\ }{{{\bf #1} {(#2)} {#3}}}}
\nc{\prc}[3]{{\it  Phys.\ Rev.\ C\ }{{\bf #1} {(#2)} {#3}}}
\nc{\prd}[3]{{\it  Phys.\ Rev.\ D\ }{{\bf #1} {(#2)} {#3}}}
\nc{\prl}[3]{{\it Phys.\ Rev.\ Lett.\ }{{\bf #1} {(#2)} {#3}}}
\nc{\pl}[3]{{\it  Phys.\ Lett.\ }{{\bf #1} {(#2)} {#3}}}
\nc{\prep}[3]{{\it Phys.\ Rep.\ }{{\bf #1} {(#2)} {#3}}}
\nc{\prsl}[3]{{\it Proc.\ R.\ Soc.\ London\ }{{\bf #1} {(#2)} {#3}}}
\nc{\ptp}[3]{{\it  Prog.\ Theor.\ Phys.\ }{{\bf #1} {(#2)} {#3}}}
\nc{\ptps}[3]{{\it  Prog\ Theor.\ Phys.\ suppl.\ }{{\bf #1} {(#2)} {#3}}}
\nc{\physa}[3]{{\it  Physica\ A\ }{{\bf #1} {(#2)} {#3}}}
\nc{\physb}[3]{{\it  Physica\ B\ }{{\bf #1} {(#2)} {#3}}}
\nc{\phys}[3]{{\it Physica\ }{{\bf #1} {(#2)} {#3}}}
\nc{\rmp}[3]{{\it  Rev.\ Mod.\ Phys.\ }{{\bf #1} {(#2)} {#3}}}
\nc{\rpp}[3]{{\it Rep.\ Prog.\ Phys.\ }{{\bf #1} {(#2)} {#3}}}
\nc{\sjnp}[3]{{\it Sov.\ J.\ Nucl.\ Phys.\ }{{\bf #1} {(#2)} {#3}}}
\nc{\spjetp}[3]{{\it Sov.\ Phys.\ JETP\ }{{\bf #1} {(#2)} {#3}}}
\nc{\yf}[3]{{\it Yad.\ Fiz.\ }{{\bf #1} {(#2)} {#3}}}
\nc{\zetp}[3]{{\it Zh.\ Eksp.\ Teor.\ Fiz.\  }{{\bf #1}  {(#2)} {#3}}}
\nc{\zp}[3]{{\it Z.\ Phys.\ }{{\bf #1} {(#2)} {#3}}}
\nc{\ibid}[3]{{\sl ibid.\ }{{\bf #1} {#2} {#3}}}
\nc{\rf}[1]{(\ref{#1})}
\nc{\nn}{\nonumber \\*}
\nc{\bfB}{\bf{B}}
\nc{\bfv}{\bf{v}}
\nc{\bfx}{\bf{x}}
\nc{\bfy}{\bf{y}}
\nc{\vx}{\vec{x}}
\nc{\vy}{\vec{y}}
\nc{\oB}{\overline{B}}
\nc{\oI}{\overline{I}}
\nc{\oR}{\overline{R}}
\nc{\rar}{\rightarrow}
\nc{\ti}{\times}
\nc{\slsh}{\hskip-5pt/}
\nc{\sm}{Standard~Model~}
\nc{\MP}{M_{\rm Pl}}
\nc{\tp}{t_{\rm Pl}}
\nc{\ave}{\bar{E}}

\nc{\eff}{{\rm eff}}
\nc{\kk}{\vek{k}}
\nc{\pp}{{\rm p}}
\nc{\ga}{g_{a\gamma}}
\nc{\vv}{\\}
\nc{\eee}{{\bf E}}
\nc{\bbb}{{\bf B}}
\nc{\qcd}{T_{\rm QCD}}
\nc{\G}{\rm \ G}
\def\vec#1{{\bf #1}}

\def\lae{\;^{<}_{\sim} \;} \def\gae{\; ^{>}_{\sim} \;} 

\def\ell{e^{c}LL}

\begin{document}
{\title{\vskip-2truecm{\hfill {{\small \\
	\hfill \\
	}}\vskip 1truecm}
{\bf   Right-Handed Sneutrinos as Curvatons   }}
{\author{
{\sc  John McDonald$^{1}$}\\
{\sl\small Dept. of Mathematical Sciences, University of Liverpool,
Liverpool L69 3BX, England}
}
\maketitle
\begin{abstract}
\noindent

     We consider the possibility that a right-handed sneutrino can serve as 
the source of energy density perturbations 
leading to structure formation in cosmology.
The cosmological evolution
 of a coherently oscillating condensate of right-handed sneutrinos is studied 
 for the case where reheating after inflation is due to perturbative 
inflaton decays. 
For the case of Dirac 
neutrinos, it is shown that some suppression of
Planck scale-suppressed corrections to the right-handed 
neutrino superpotential is necessary in order 
to have sufficiently late decay of the right-handed 
sneutrinos. $cH^{2}$ corrections to the sneutrino mass squared term 
must also be suppressed during inflation ($|c| \lae 0.1$), in which case,  
depending on the magnitude of $|c|$ during inflation, 
a significantly 
blue (if $c > 0$) or red (if $c < 0$) perturbation spectrum is possible.    
R-parity must also be broken in order 
to ensure that the Universe is not overclosed by LSPs from the late decay 
(at temperatures $1-10 \MeV$) of the right-handed sneutrino condensate. The 
resulting expansion rate during inflation can be significantly smaller
 than in conventional supersymmetric inflation models (as low as 
$10^{6} \GeV$ is possible). 
For the case of Majorana neutrinos, a 
more severe suppression of Planck-suppressed 
superpotential corrections is required. In addition,   
the Majorana sneutrino condensate is likely to be thermalised before 
it can dominate the energy density, which would exclude
the Majorana right-handed sneutrino as a curvaton.

\end{abstract} 
\vfil
 \footnoterule {\small $^1$mcdonald@amtp.liv.ac.uk}   
 \newpage 
\setcounter{page}{1}                   

 \section{Introduction}  

      The observation of neutrino masses and mass splittings, via solar and atmospheric  
neutrinos \cite{san}, strongly suggests the existence of  
right-handed (r.h.) neutrinos. In extensions of the 
Minimal Supersymmetric Standard Model (MSSM) \cite{mssm} 
which can accomodate neutrino masses we therefore 
expect to have right-handed sneutrinos. In the cosmology of the MSSM and 
its extensions, Bose 
condensates of scalar  fields such as squarks and sleptons form naturally after inflation 
\cite{em}. These may have important consequences for cosmology; for example, 
they allow for the 
possibility of baryogenesis and leptogenesis via the Affleck-Dine mechanism 
\cite{em,ad}. Since right-handed sneutrinos may also form condensates, it is important 
to consider in some detail the cosmological evolution of a r.h. sneutrino condensate.   

         Due to the weak coupling of the r.h. sneutrinos 
to the MSSM fields, a condensate of r.h. sneutrinos will be 
long-lived and so may come to dominate the energy density of the Universe 
before it decays.  The question of whether the r.h.
sneutrino condensate can dominate the energy 
density of the Universe when it decays has recently acquired some importance. It has 
been noted \cite{moll,es} that if a scalar dominates the energy density when it decays, 
and if 
that  scalar is effectively massless during inflation, then quantum fluctuations of the 
scalar during inflation can in principle account for the primordial 
energy density perturbations leading to structure formation 
\cite{moll,es,moroi}. This has been labelled the curvaton scenario \cite{lw,liddle}. 
Should the curvaton be able to account for the density perturbations, 
the parameters of the  inflation model would be more weakly constrained than in the 
conventional case where density perturbations arise from quantum fluctuations of the inflaton. 

                 Thus if there exists a 
natural curvaton candidate such as the r.h.
sneutrino, it is important to confirm or 
exclude that candidate as a curvaton\footnote{Models have 
recently been proposed where the curvaton corresponds 
to an MSSM flat direction scalar \cite{enmz} and an MSSM Higgs scalar \cite{gil}.}.  
The main goal of this paper is to 
investigate the possibility of a r.h. sneutrino curvaton.

     The masses of the r.h. sneutrinos and their coupling to the MSSM fields are  
determined by the model of neutrino masses, in  
particular whether they have Dirac masses or Majorana 
masses via a see-saw mechanism \cite{san}.  The Yukawa coupling of the 
r.h. sneutrino to the MSSM fields plays a 
fundamental role in determining the evolution of the r.h. sneutrino condensate,  in 
particular its decay temperature, its effective mass from interacting with the 
background of inflaton decay  products \cite{eall} and its
 rate of thermalisation/scattering from the background. 

    Throughout this paper we will consider the simplest model for 
inflation and reheating, coresponding to a constant expansion rate during inflation 
followed by formation of a coherently 
oscillating  inflaton condensate and reheating due to perturbative inflaton decays.    

        The paper is organised as follows. In Section 2 we discuss the cosmological  
enviroment due to perturbative inflaton decays. In Section 3 we consider the evolution 
of the r.h. sneutrino condensate in this enviroment. In Section 4 we 
present our conclusions.  

\section{Cosmological Background from Perturbative Inflaton Decays}

          After inflation we consider the inflaton $S$ to have a mass $m_{S}$ and  to be 
coherently oscillating  about the minimum of its potential. The inflaton is assumed to 
decay  into pairs of relativistic MSSM particles with initial energy $m_{S}/2$. If the 
initial energy of the decay products is sufficiently 
large, the initial scattering rate of the inflaton decay products may be small enough   
that thermalisation only occurs once the Universe has expanded sufficiently for the 
decay product scattering rate  $\Gamma_{sc}$ to exceed the expansion rate $H$. 
There are therefore two 
possibilities: 
\newline {\bf (a) Instantaneous Thermalisation.} The inflaton decay products 
thermalise  immediately after decay. In this case there will be two epochs: (i) Inflaton 
Matter  Domination (IMD), where the energy density of the  Universe is dominated by 
the coherently oscillating inflaton, and (ii) Radiation Domination (RD), 
defined to mean domination of the energy density  of the Universe  by 
relativistic inflaton decay products, not necessarily thermalised. 
 \newpage {\bf (b) Non-instantaneous Thermalisation.} The relativistic inflaton decay  
products are unthermalised initially. In this case we will show  that thermalisation 
cannot occur during IMD and so must occur during RD.  Therefore in this case the 
background will have three distinct epochs:  (i) IMD, (ii) RD pre-thermalisation and 
(iii) RD post-thermalisation.   

\subsection{Inflaton decay product thermalisation}      

         Prior to thermalisation there is a spectrum of decay products, ranging from
  red-shifted products from the earliest decays (occuring at the end of inflation) to 
products  from the most recent decays. Most of the energy density in decay products 
will come 
from the most recent decays. Following the discussion of \cite{jtherm}, the spectrum 
of unthermalised decay products as a function of energy during IMD and 
RD epochs is given by 
 \be{jt1} \frac{dn}{dE} \approx \frac{3}{2}  
\left(\frac{H}{H_{R}}\right)^{\gamma} \frac{\rho_{S}\left(H_{R}\right)}{m_{S}}  
\frac{E^{1/2}}{m_{S}^{3/2}}   ~,\ee where $\gamma = 1$ during IMD ($3/4$ during 
RD), $n(E)$ is the number density of decay products with energy less than $E$ 
and $H_{R}$ is the expansion 
rate at  the onset of radiation domination. Thus $n(E) \propto E^{3/2}$
 and so 80$\%$ of the decay products at a given  time have energy between $E_{d}$ 
and $E_{d}/3$, where $E_{d}$  is the energy of the most recent decay 
products. In addition, once thermalisation by scattering begins,  the lower energy decay 
products tend first to increase  their energy by scattering from the more numerous  
higher-energy decay products in the spectrum  \cite{sd,allah}, so that they may be regarded 
as higher-energy decay products as far as thermalisation is concerned.  Thus we will 
consider the energy of the decay products at a given time to be approximately  $E_{d}$. 
During IMD, $E_{d} \approx m_{S}/2$.  Once the inflaton condensate has decayed 
away and the  Universe enters the RD epoch, the energy of the dominant unthermalised 
decay products will be red-shifted to  $ E_{d} \approx \left(\frac{a_{R}}{a}\right) 
\frac{m_{S}}{2}$, where $a_{R}$ is 
the scale  factor at the onset of radiation domination.                

    The centre of mass (CM) cross-section of the  relativistic inflaton decay products is 
 \be{e1} \sigma_{sc}  \approx \frac{\alpha_{sc}^{2}}{E_{CM}^{2}} ~,\ee
where $E_{CM} \approx 2E_{d}$ and  $\alpha_{sc} = g^{2}/4\pi$, where $g$ is a 
typical MSSM gauge or Yukawa coupling.  Therefore the scattering rate of the inflaton 
decay products is  $\Gamma_{sc} = n \sigma_{sc}$, where $n$ is the number density 
of inflaton decay products.   

                    During IMD most of the decay products at a given scale factor are 
produced during the previous e-folding.   The  
energy density in the inflaton condensate during IMD is  
\be{e3} \rho_{S} = \left(\frac{a_{e}}{a}\right)^{3}e^{-\Gamma_{d}t}\rho_{I}   
~,\ee 
where $\Gamma_{d}$ is the decay rate of the inflaton, $a_{e}$ is the scale factor at 
the  end of inflation and $\rho_{I}$ is the energy density during inflation, assumed 
constant.  Therefore the number of inflatons which decay during an e-folding ($\delta t 
\approx H^{-1}$)  (and so the number of inflaton decay products produced) is  
\be{e4} n \approx   
\frac{\Gamma_{d}}{H}\frac{\rho_{S}}{m_{S}}     ~\ee
 for $\Gamma_{d}/H < 1$. 
The condition for thermalisation to occur during IMD, $\Gamma_{sc} 
\gae  H$, is then 
\be{e5} \frac{3 \alpha_{sc}^{2} M^{2} \Gamma_{d}}{m_{S}^{3}} \gae 1    ~,\ee 
where $M =M_{Pl}/\sqrt{8\pi}$ and we have used $E_{d} = m_{S}/2$ and $\rho_{S} \approx 
3 M^{2} H^{2}$ during IMD. \eq{e5} is {\it independent} of the scale 
factor, so if it is not 
satisfied immediately at the  end of inflation, it will not be satisfied during IMD. 
Therefore for the case of  perturbative inflaton decay, thermalisation must either be  
instantaneous or must occur during RD.       

   After IMD, the number density and energy of the relativistic decay products is  
\be{e6}  n = \left(\frac{a_{R}}{a}\right)^{3} 
n(a_{R}) \;\;\; ; \;\;  E_{d} \approx \left(\frac{a_{R}}{a}\right)\frac{m_{S}}{2}     
~.\ee 
The inflaton decay rate, 
$\Gamma_{d}$, may be expressed in terms  of the reheating 
temperature\footnote{The reheating 
temperature is defined in the following to be the  
temperature of a thermalised Universe at the onset of 
radiation domination. It is therefore used generally to parameterise the 
energy density at the onset of radiation domination, even if 
the relativistic decay products at that time have not yet thermalised.}, 
$T_{R}$, as  $\Gamma_{d} = k_{T_{R}}T_{R}^{2}/M_{Pl}$,  
where $k_{T_{R}} = (4 \pi^{3} g(T_{R})/45)^{1/2}$ and $g(T_{R})$  is the 
effective number of massless degrees of freedom in thermal equilibrium \cite{tk}. 
(In the 
following we will consider $k_{T_{R}} \approx 20$, corresponding to the 
field content of the MSSM with $g(T_{R}) \approx 200$.) 

The thermalisation condition $n \sigma_{sc} \gae H$  (where $H =  
\left(\frac{a_{R}}{a}\right)^{2} H(a_{R})$ during RD)  
then implies that  thermalisation occurs 
at scale factor $a_{th}$ given by 
\be{e7} \frac{a_{th}}{a_{e}}  \approx  \frac{m_{S}^{3}}{3 \alpha_{sc}^{2} M^{2} 
\Gamma_{d}}  \left(\frac{H_{I}}{\Gamma_{d}}\right)^{2/3}    ~,\ee 
with the corresponding temperature given by
\be{e7a} T_{th} \approx  \frac{3}{\sqrt{8 \pi}} \frac{k_{T_{R}}\alpha_{sc}^{2} 
M T_{R}^{3} }{m_{S}^{3}}    ~.\ee
Here $H_{I}$ is the expansion rate during inflation.

  \subsection{Scalar field squared expectation value of the inflaton decay product background}    

    When discussing the evolution of the r.h.
sneutrino condensate,  we will need the expectation value 
$<\Psi^{2}>$, where $\Psi$ represents a generic real MSSM scalar in the inflaton 
decay product background \cite{eall}.  
\newline {\bf (i) Unthermalised Decay Products during IMD.}  
If we consider the average momentum of the scalar modes in the inflaton 
decay product background to be $\sim k$, then
 the average energy density of a real massless scalar field $\Psi$ is  
\be{e7b} <\rho_{\Psi}> = <\frac{1}{2} \dot{\Psi}^{2} + \frac{1}{2} (\nabla 
\Psi)^{2}  > \approx k^{2} <\Psi^{2}>    ~.\ee 
Thus with $k \approx E_{d} \approx 
m_{S}/2$ and $<\rho_{\Psi}> = 
f_{\Psi}\rho_{d}$,  where  $\rho_{d} \approx (\Gamma_{d}/H) \rho_{S}$ is the energy 
density of the inflaton decay 
products during IMD  and $f_{\Psi}$ is the fraction of the total inflaton decay product energy 
density in the 
real scalar field $\Psi$,  we obtain  
\be{e9} <\Psi^{2}> \approx \frac{f_{\Psi}\rho_{d}}{k^{2}} 
 \approx \frac{12 f_{\Psi} \Gamma_{d} M^{2}H}{m_{S}^{2}}   ~.\ee 
\newline {\bf (ii) Unthermalised Decay Products during RD.}  In this case the energy
 of the dominant decay products red-shifts  as $E_{d} \approx 
\left(\frac{a_{R}}{a}\right) \frac{m_{S}}{2}$  whilst 
$\rho_{d} \propto a^{-4}$ for relativistic decay products, 
with $\rho_{d}(a_{R}) \approx \rho_{S}(a_{R})$. 
Thus  $<\Psi^{2}> \propto \rho_{d}/E_{d}^{2} 
\propto a^{-2} \propto H$. Therefore the same relation, \eq{e9}, between 
$<\Psi^{2}>$ and $H$ also holds during RD. 
 \newline {\bf (iii) Thermalised Decay Products}.  In this case the energy density of the 
inflaton decay products is  $\rho_{d} = \frac{\pi^{2}g(T)T^{4}}{30}$, where for a real  
scalar field $g(T) = 1$. Therefore, with  $k \approx T$ for 
effectively massless thermalised particles, we find 
 \be{e10}  <\Psi^{2}>  \approx  \gamma_{T} 
T^{2} \;\;\; ; \;\;  \gamma_{T} =  \frac{\pi^{2}}{30}     ~.\ee 
We note that in the case where the inflaton decay  products thermalise 
immediately at the end of  
inflation,  the temperature during IMD is related to $H$ by \cite{tk} 
\be{e11}  T = k_{r} (M_{Pl}HT_{R}^{2})^{1/4}  \;\;\; ; \;\;\;  k_{r} = 
\left(\frac{9}{5 \pi^{3} g(T)}\right)^{1/8}   ~.\ee  

\section{Right-Handed Sneutrino Condensate Evolution}  

\subsection{Neutrino masses and the r.h. sneutrino scalar potential}

      For simplicity we will consider a single neutrino generation.  
The superpotential of the r.h. neutrino superfield, $N$, is given by 
\be{e12}   W_{\nu} = \lambda_{\nu} N H_{u} L 
+ \frac{M_{N} N^{2}}{2}    ~,\ee
where $H_{u}$ and $L$ are the MSSM Higgs and charged 
lepton superfield $SU(2)_{L}$ doublets \cite{mssm}. 
The corresponding scalar potential for the r.h. 
sneutrino is then $V(N) = \frac{m_{\tilde{N}}^{2}}{2}N^{2}$ 
(with $N$ a conventionally normalised real scalar field),
where $m_{\tilde{N}}^{2} = m_{o}^{2} + M_{N}^{2} + m_{eff}^{2}$. Here 
$m_{o}^{2}$ is the conventional SUSY breaking mass squared term ($m_{o} \sim 100 \GeV$) 
whilst $m_{eff}^{2} \equiv \lambda_{\nu}^{2}<\Psi^{2}>$ is the effective mass squared 
term due to the 
interaction of the r.h. sneutrinos with the inflaton decay product backround \cite{eall}. 
In addition, we expect terms due to 
Planck-scale suppressed interactions. For now we will consider 
the evolution of the sneutrino condensate in the absence of such terms. 

    If $M_{N} = 0$ we will have Dirac neutrino masses, with 
$m_{\nu} = \lambda_{\nu} v_{u}$. The r.h. sneutrino mass is then given by 
$m_{\tilde{N}}^{2} = m_{o}^{2} + m_{eff}^{2}$. 
If $M_{N} \gg \lambda_{\nu}v_{u}$ (where $v_{u}= <H_{u}>$) we will have 
Majorana  neutrino masses from the see-saw mechanism, $m_{\nu} = 
\frac{\lambda_{\nu}^{2} v_{u}^{2}}{M_{N}} $,
such that the Yukawa coupling $\lambda_{\nu}$ can 
be expressed as a function of $m_{\nu}$, 
\be{13a} \lambda_{\nu} = \left(\frac{m_{\nu}M_{N}}{v_{u}^{2}}\right)^{1/2}   
~.\ee
The usual idea of the see-saw mechanism \cite{san} 
is to consider the magnitude of $\lambda_{\nu}$ to be similar to the 
charged lepton Yukawa couplings, which requires e.g.  $M_{N} \approx 10^{9} \GeV$
 for $\lambda_{\nu} \approx \lambda_{\tau} \approx 10^{-2}$ and $m_{\nu} \approx 0.1$eV. 
However, other natural 
mass scales could also be of interest, for example $M_{N} \sim 100 \GeV - 1 \TeV$,
as suggested by the electroweak scale and by the scale of the SUSY mass term $\mu H_{u}H_{d}$ 
of the MSSM superpotential \cite{mssm}.
 
\subsection{Conditions for r.h. sneutrino to act as a curvaton}

          R.h. sneutrino oscillations begin once the r.h. 
sneutrino mass satisfies $m_{\tilde{N}} \gae H$. 
We denote the scale factor at this time by $a_{osc}$ and the homogeneous sneutrino 
expectation value by $N_{osc}$. During inflation, in order to serve as a curvaton, the 
r.h. sneutrino must be effectively massless, $m_{\tilde{N}} \ll H_{I}$. The quantum 
fluctuation of 
an effectively massless r.h. sneutrino mode at horizon crossing is given by $\delta N 
\approx H_{I}/2\pi$ \cite{tk}. In the case of a scalar potential consisting 
purely of a mass term ($\propto N^{2}$),
 once the perturbation mode is stretched outside the horizon by the expansion of the Universe,  
its amplitude on sub-horizon scales will evolve in the 
same way as the homogeneous field. This can be seen by considering $N = N_{o} + \delta N$, where
$N_{o}$ is the homogeneous field and $\delta N$ is a perturbation of wavenumber $\vec{k}$. The 
equations of motion for these are
\be{sf1} \ddot{N}_{o} + 3 H \dot{N}_{o} = -V^{'}(N_{o})   ~\ee
and 
\be{sf2} \delta \ddot{N} + 3 H \delta \dot{N} -\frac{\vec{k}^{2}}{a^{2}}\delta N  
= -V^{''}(N_{o})\delta N  ~.\ee
For a mode outside the horizon, the $\vec{k}^{2}/a^{2}$ term $(\ll H^{2})$ 
will effectively play no role in the evolution of the scalar field. Therefore, for 
$V(N)  \propto N^{2}$, \eq{sf1} $\leftrightarrow$ \eq{sf2} under 
$N_{o} \leftrightarrow \delta N$. Thus $N_{o}$ and $\delta N$ will evolve in the 
same way. Therefore $\delta N/N$ for a superhorizon perturbation will be fixed by its 
value at the onset of oscillations, $(\delta N/N)_{osc}$.  
Once coherent oscillations of the r.h. sneutrino begin,
the energy density in the r.h. sneutrino field will be proportional to its amplitude 
squared. Therefore if the energy density
 of the Universe becomes dominated by the r.h. sneutrino oscillations before 
the sneutrinos decay, the energy density perturbation when a given mode re-enters the 
horizon will be given by 
\be{e14} \delta_{\rho} \equiv \frac{\delta \rho}{\rho} \approx \left(\frac{2 \delta 
N}{N}\right)_{osc}  
= \frac{H_{I}}{\pi N_{I}}     ~.\ee
(A more precise calculation gives the same result up to a 
factor of the order of 1 \cite{lw}.) For $H_{I}$ and $N_{I}$ constant this 
corresponds to a scale-invariant perturbation spectrum. 
In order to account for the observed CMB 
temperature fluctuations, we then require that $\delta_{\rho} 
\approx 10^{-5}$ \cite{specobs}.

\subsection{Condensate evolution without Planck-suppressed terms}

        The evolution of the r.h. sneutrino expectation value depends on 
the inflaton decay product background and reheating temperature. We will 
consider the case where the r.h. sneutrino begins coherent oscillations during the IMD 
epoch, 
$a_{osc} < a_{R}$. Oscillations begin at $H_{osc} \approx m_{\tilde{N}} \gae 100 \GeV$. 
The condition $H_{R} < H_{osc} \approx m_{\tilde{N}}$ then implies an upper bound on $T_{R}$, 
\be{e15}   T_{R} \lae \left(\frac{m_{\tilde{N}} M_{Pl}}{k_{T_{R}}}\right)^{1/2}   
\approx 7.8 \times 10^{9} \left(\frac{m_{\tilde{N}}}{100 \GeV} \right)^{1/2}  \GeV  ~.\ee
In particular, if $T_{R} \lae 10^{8} \GeV$, as would be required by the thermal 
gravitino upper bound \cite{tg,sark}
 {\it if} the inflaton decay products are thermalised at the onset of 
radiation domination, then r.h. sneutrino oscillations would generally begin during IMD. 
However, it is possible that thermalisation of the relativistic decay 
products could occur at $T_{th} \lae 10^{8} \GeV$ even though $T_{R} > 
10^{8}\GeV$, in which case the thermal gravitino bound on 
$T_{R}$ could be evaded. 

    We will also assume that thermalisation of the inflaton decay products occurs 
after the Universe becomes radiation dominated. This is true if \eq{e5} is {\it not} 
satisfied, which in turn requires that the inflaton mass satisfies
\be{e16} m_{S} \gae 6 \times 10^{11} \alpha_{sc}^{2/3} \left(\frac{T_{R}}{10^{8} 
\GeV}\right)^{2/3} \GeV   ~,\ee
where $\alpha_{sc}$, being due to typical MSSM couplings, 
is not expected to be much smaller than 1.
We finally assume that $m_{\tilde{N}}^{2}$ is dominated by the time-independent 
terms when the sneutrino oscillations begin,
 $m_{\tilde{N}}^{2} \approx m_{\tilde{N}_{c}}^{2} \equiv m_{o}^{2} + M_{N}^{2}$.
This requires that $m_{eff}^{2} < m_{\tilde{N}_{c}}^{2}$ 
at $H_{osc} \approx m_{\tilde{N}_{c}}$. Using 
$<\Psi^{2}>$ for an unthermalised background during IMD, \eq{e9}, this requires that 
\be{e17} m_{S}^{2} \gae  
\frac{12 k_{T_{R}} f_{\Psi} \lambda_{\nu}^{2}  
M T_{R}^{2}}{\sqrt{8 \pi} m_{\tilde{N}_{c}} } 
~.\ee
For the case of Dirac neutrinos this implies that 
\be{e17a} m_{S} \gae 1.1 \times 10^{5} f_{\Psi}^{1/2} 
\left(\frac{100 \GeV}{m_{\tilde{N}_{c}}}\right)^{1/2}
\left(\frac{T_{R}}{10^{8}\GeV}\right)
\left(\frac{m_{\nu}}{0.1 \eV}\right) \GeV  ~,\ee
where we have assumed $v_{u} \approx 100 \GeV$, 
whilst for the case of Majorana neutrino masses
\be{e18} m_{S} \gae 1 \times 10^{11} 
f_{\Psi}^{1/2}
\left(\frac{M_{N}}{m_{\tilde{N}_{c}}}\right)^{1/2} 
\left(\frac{T_{R}}{10^{8} \GeV}\right)
\left(\frac{m_{\nu}}{0.1 \eV}\right)^{1/2} 
\GeV ~.\ee
This assumption will give the largest r.h. sneutrino energy density at late times 
for a given $T_{R}$ and $N_{osc}$, since the effect of having $m_{eff} > m_{\tilde{N}_{c}}$
at $H \approx m_{\tilde{N}_{c}}$ would be to cause r.h. sneutrino 
oscillations to begin earlier and so to 
experience a greater dilution of the r.h. sneutrino energy density due to expansion. 
From now on we will consider $m_{\tilde{N}} \approx m_{\tilde{N}_{c}}$.  

    A fundamental condition for the r.h. sneutrino to play the role of a curvaton is 
that the 
r.h. sneutrinos decay after the Universe becomes dominated by the energy density in their
coherent oscillations. The r.h. sneutrino decay rate is given by 
\be{e19} \Gamma_{\tilde{N}d} \approx 
\frac{\lambda_{\nu}^{2} m_{\tilde{N}}}{4 \pi}      ~.\ee
The r.h. sneutrinos decay once $\Gamma_{\tilde{N}d} \approx H$. 
At the onset of r.h. sneutrino oscillations, 
$\rho_{\tilde{N}} \approx m_{\tilde{N}}^{2}N_{osc}^{2}/2$ 
whilst $\rho_{S} \approx 3 H_{osc}^{2}M^{2} \approx 3  m_{\tilde{N}}^{2}  M^{2}$.  
During IMD, the energy density in the inflaton oscillations and the r.h.  sneutrino 
oscillations are both evolve as $a^{-3}$. Thus $\rho_{\tilde{N}}/\rho_{S}$ is constant.  
Once the Universe is radiation dominated, the 
energy density in the dominant relativistic background, $\rho$, evolves as $a^{-4}$ 
whilst the energy density in the r.h. sneutrino evolves as $a^{-3}$. Therefore once $a 
> a_{R}$, 
\be{e21} \frac{\rho_{\tilde{N}}}{\rho}
 = \left(\frac{a}{a_{R}}\right) 
\left(\frac{\rho_{N}}{\rho_{S}}\right)_{osc} \approx 
\left(\frac{a}{a_{R}}\right)\frac{N_{osc}^{2}}{6 M^{2}} ~.\ee
Thus the Universe becomes r.h. sneutrino dominated once $H < H_{dom}$, where
\be{e22} H_{dom} \approx \left(\frac{N_{osc}^{2}}{6 M^{2}}\right)^{2} H_{R}    ~.\ee
The condition that r.h. sneutrino decay occurs after r.h. sneutrino domination, 
$\Gamma_{\tilde{N}d} < H_{dom}$, then implies that 
\be{e22a} m_{\tilde{N}} \lae
\left(\frac{4 \pi k_{T_{R}}T_{R}^{2}}{ \lambda_{\nu}^{2} M_{Pl}}\right) 
\left(\frac{N_{osc}^{2}}{6 M^{2}}\right)^{2}    ~.\ee
We will refer to \eq{e22a} as the late decay condition in the following.

   For the case of Dirac neutrino masses the late decay condition becomes
\be{e22b} m_{\tilde{N}} \lae \left(\frac{4\pi v_{u}^{2}}{m_{\nu}^{2}}\right)
\left(\frac{k_{T_{R}}T_{R}^{2}}{M_{Pl}}\right) 
\left(\frac{N_{osc}^{2}}{6 M^{2}}\right)^{2}    ~,\ee
which implies that
\be{e22c} m_{\tilde{N}} 
\lae 5.8 \times 10^{21} \left(\frac{0.1 \eV}{m_{\nu}}\right)^{2}
 \left(\frac{T_{R}}{10^{8} \GeV}\right)^{2}
 \left(\frac{N_{osc}}{M}\right)^{4} \GeV  ~.\ee
Thus in order to have $m_{\tilde{N}} \gae m_{o} \approx 100 \GeV$ we require that 
$N_{osc}/M \gae 1 \times 10^{-5}$. We will see that values of $N_{osc}/M$ in this range require 
some suppression of Planck-scale suppressed 
non-renormalisable corrections to the r.h. neutrino superpotential. 

   For the case of Majorana neutrino masses the late decay condition becomes
\be{e23} m_{\tilde{N}} \lae \left(\frac{m_{\tilde{N}}}{M_{N}}\right)^{1/2}
 \left(\frac{4\pi v_{u}^{2}}{m_{\nu}}\right)^{1/2} 
\left(\frac{k_{T_{R}}T_{R}^{2}}{M_{Pl}}\right)^{1/2} 
\frac{N_{osc}^{2}}{6 M^{2}}    ~,\ee
such that
\be{e24} m_{\tilde{N}} \lae 7.6 \times 10^{5} 
 \left(\frac{m_{\tilde{N}}}{M_{N}}\right)^{1/2}
\left(\frac{0.1 \eV}{m_{\nu}}\right)^{1/2} 
\left(\frac{T_{R}}{10^{8} \GeV}\right) 
\left(\frac{N_{osc}}{M}\right)^{2} \GeV    ~.\ee
We will 
restrict attention to r.h. neutrino masses 
$M_{N} \gae m_{o} \approx 100 \GeV$, since masses smaller than 
this would have little motivation from neutrino mass models or natural particle 
physics scales. In this case 
$m_{\tilde{N}} \approx M_{N}$. \eq{e24} then shows that a right-handed 
Majorana sneutrino curvaton with $ M_{N} \gae 100\GeV$ is possible only
if $N_{osc}$ is not very small compared with the reduced Planck scale $M$.
($N_{osc}/M \gae 0.01$ if $T_{R} \lae 10^{8} \GeV$).  
It also shows that 
$T_{R}$ cannot be very small compared with the thermal gravitino upper bound 
of $10^{8} \GeV$ if $N_{osc} \lae M$. The requirement that 
$N_{osc}$ is not very small compared with $M$
is a strong constraint on a Majorana r.h. sneutrino 
curvaton, since, as we discuss below, it requires a high degree of 
suppression of Planck-scale 
suppressed contributions to the r.h. sneutrino superpotential.

       From \eq{e14} and the density perturbation constraint,
$\delta_{\rho} \approx 10^{-5}$, the expansion rate during inflation is  
$H_{I} = \pi \delta_{\rho} N_{I} \gae 10^{9} \GeV$ if 
$N_{I} \approx N_{osc} \gae 1 \times 10^{-5}M$, corresponding to the case of Dirac
neutrino masses. This can be substantially lower than
 the typical value of the expansion rate during inflation found 
in conventional SUSY inflation models, $H_{I} \approx 10^{13} \GeV$ \cite{dti,fti,chaotic}. 
For the case of Majorana neutrino masses with $N_{I} \approx N_{osc} \gae 0.01M$, 
the corresponding bound on $H_{I}$ is $H_{I} 
\gae 7 \times 10^{11} \GeV$. These bounds assume that 
$N$ does not evolve significantly from the end of inflation 
to the onset of sneutrino oscillations, 
so that $N_{osc} \approx N_{I}$. 
We will see that the range of allowed $H_{I}$ can be increased 
if this assumption is altered.
 
\subsection{Effect of Planck-suppressed terms}

               In general, in addition to the 
globally SUSY scalar potential we expect (in the absence of specific symmetries)  
terms suppressed by powers of the reduced Planck mass, $M = M_{Pl}/\sqrt{8 \pi}$, 
corresponding to the natural scale of supergravity (SUGRA) corrections \cite{drt,h2}. 
Thus we expect 
Planck-scale suppressed non-renormalisable terms to appear in the r.h. neutrino 
superpotential. 
We also expect contributions to the mass squared term of the form $cH^{2}$, where $|c|$
 is model-dependent but expected to be of the order of 1 in the simplest models. These 
arise from terms in the full Lagrangian of the form $\frac{1}{M^{2}} \int d^{4}\theta 
S^{\dagger}S N^{\dagger}N = |F_{S}|^{2}/M^{2}$, where 
$S$ is the inflaton field (or any other field with a non-zero F-term contributing to the 
energy density of the Universe). 
\newline {\bf (i) $cH^{2}$ corrections}                                      

      During inflation the value of $|c|$ is constrained by the deviation of the curvaton 
perturbation from scale-invariance. We consider
 the r.h. sneutrino potential during inflation to be
\be{e27}    V(N) \approx \frac{1}{2} c H^{2} N^{2}    ~.\ee
The expansion rate during inflation is then
\be{e28} H_{I}^{2}  = \frac{1}{3 M^{2}} \left(\rho_{S} + \frac{c H_{I}^{2} N^{2}}{2} 
\right) ~\ee
where $\rho_{S}$ is the energy density of the inflaton field, which we assume to be
 constant. 

The index $n$ of the perturbation spectrum as a function of present wavenumber $k$ 
is given by 
\be{e29} n = 1 + \frac{2 k}{\delta_{\rho}}
\frac{d\delta_{\rho}}{dk}     ~,\ee
such that $\delta \rho/\rho \propto k^{\frac{n-1}{2}}$ and 
$n=1$ corresponds to scale-invariance. 
$\delta_{\rho}$ will remain constant once the perturbation is outside the horizon, 
since both $N$ and $\delta N$ evolve in the same way for $V(N) \propto N^{2}$. 
Therefore we have 
\be{e30} \delta_{\rho} \equiv \frac{2 \delta N}{N} = 
\left(\frac{H_{I}}{\pi N} \right)_{a_{\lambda}}   ~,\ee
where $a_{\lambda}$ is the scale-factor at which a perturbation of wavelength 
$\lambda$ exits the horizon.  

       The value of $N$ as a function of the scale factor 
during inflation is given by the solution of 
\be{e31} \ddot{N} + 3 H \dot{N} = -c H^{2} N   ~\ee
With $H \propto a^{-m}$, this has a solution of the form $N \propto a^{\gamma}$, 
where
\be{e32} \gamma = \frac{1}{2} \left[ -(3-m) + \sqrt{(3-m)^{2} - 4 c} \right]   ~.\ee
Thus during inflation ($m=0$) the solution corresponds to 
\be{e33} \gamma = \frac{1}{2} \left[ -3 + \sqrt{9 - 4 c} \right]  ~.\ee
If $|4c| \ll 9$ then during inflation $\gamma \approx -c/3$ and 
so $N/N_{\lambda} =  
(a/a_{\lambda})^{-c/3}$, where $N_{\lambda}$ is the value at $a_{\lambda}$. 
Thus for a given $N$ and $a$ (for example, their values at 
the end of inflation), 
$N_{\lambda} \propto a_{\lambda}^{-c/3}$.   
The wavenumber at present is related to the scale factor at horizon exit
($\lambda \approx H_{I}^{-1}$) by $k = 2 \pi H_{I} a_{\lambda}/a_{p}$, where 
$a_{p}$ is the scale factor at present. Thus 
\be{e34} \frac{d N_{\lambda}}{dk} 
= \frac{dN_{\lambda}}{d a_{\lambda}} \frac{d a_{\lambda}}{d k} 
= - \frac{cN_{\lambda}}{3k}    ~.\ee   
For the case where the energy density is dominated by $\rho_{S}$, we have
\be{e35}   \frac{d \delta_{\rho}}{d N_{\lambda}} 
\approx - \frac{H}{\pi N_{\lambda}^{2}}     ~.\ee
Therefore 
\be{e36} n = 1 + \frac{2k}{\delta_{\rho}} \frac{d \delta_{\rho}}{d N_{\lambda}}
\frac{d N_{\lambda}}{d k}  \approx   1 + \frac{2 c}{3}    ~.\ee
Note that $c> 0$ ($<0$) results in a 
blue (red) spectrum of perturbations. Observation requires that 
$|\Delta n| < 0.1$ \cite{specobs}. Therefore during inflation we must have 
\be{e37}     |c| < 0.15 \left|\frac{\Delta n}{0.1}\right|     ~.\ee
If $|c|$ is close to this upper bound then a significant blue or red perturbation 
spectrum is expected. We note that this would allow the r.h. sneutrino curvaton scenario 
to be consistent with the recent observation by WMAP of a blue perturbation spectrum 
on comoving scales of the order of 500 Mpc, with $n = 1.10^{+0.07}_{-0.06}$ \cite{wmap}.

     Thus $|c|$ is constrained to be not much larger than 0.1 during inflation.
 This is significantly larger than the value $|c| \sim 1$ expected on dimensional grounds 
in SUGRA models \cite{eta}. 
This problem is similar to the conventional '$\eta$-problem' encountered in SUSY 
inflation 
models \cite{dti,fti,chaotic}, where order $H^{2}$ corrections lead both to a large deviation from
 scale-invariance and to insufficient 
slow-roll inflation \cite{eta}. It has a natural solution 
in the case of SUSY D-term hybrid inflation (driven by the energy density of
a Fayet-Illiopoulos D-term) \cite{dti}, in which case $|F| = 0$ and so $|c| = 0$ during inflation 
(although a non-zero value is expected once inflation ends and coherent inflaton
 oscillations begin, since $\int d^{4}\theta 
S^{\dagger}S \Phi^{\dagger}\Phi = \left( \left| \partial_{\mu} 
S \right|^{2} + |F_{S}|^{2} + ... \right) |\Phi|^{2}$, with $\Phi$ a general scalar 
superfield).
 Since in the r.h. 
sneutrino curvaton scenario inflation must still be driven by an inflaton, 
it is possible that the 
mechanism which suppresses $|c|$ for the inflaton during 
inflation also suppresses $|c|$ for the curvaton. 

   After inflation we generally expect a non-zero F-term from the energy density of the
 coherently oscillating inflaton field. The solution 
of \eq{e31} during IMD corresponds to \eq{e32} with $m = 3/2$. 
Thus assuming that $N$ has a constant value during inflation, $N_{I}$, 
we find
\be{e38} \frac{N_{osc}}{N_{I}} = \left(\frac{a_{osc}}{a_{e}}\right)^{\gamma} = 
\left(\frac{H_{I}}{H_{osc}}\right)^{2 \gamma/3}      ~,\ee
where $H_{osc} \approx m_{\tilde{N}}$. 

 We previously derived lower bounds on the value of $H_{I}$ compatible with 
the observed density perturbations 
based on the assumption that $N_{osc} \approx N_{I}$. 
However, if $c < 0$ during IMD then the growth of $N$ 
after inflation will allow a wider 
range of $H_{I}$ to be compatible with a given value of $N_{osc}$, 
\be{e39} \frac{2\gamma}{3} = \frac{{\rm ln}\left(\frac{\pi \delta_{\rho} 
N_{osc}}{H_{I}}\right)}{
{\rm ln}\left( \frac{H_{I}}{m_{\tilde{N}}} \right)}    ~,\ee
where we have used $N_{I} \approx H_{I}/\pi\delta_{\rho}$ in \eq{e38}.
For example, for the case of Dirac neutrino masses, 
with $\delta_{\rho} \approx 10^{-5}$, $m_{\tilde{N}} \approx 100 \GeV$ 
and $N_{osc} \approx 10^{-4} M$, it is possible to have $H_{I} 
\approx 10^{6} \GeV$ if $\gamma = 1.45$, corresponding to $c = -4.3$. 
This also shows that $|c|$ need not be small compared with 1 after inflation.

      We conclude that the $cH^{2}$ correction to the r.h. sneutrino mass squared term 
is no more problematical 
for the curvaton than for the conventional inflaton: we require that $|c| \lae 0.1$ during 
inflation, in which case a significantly blue or red perturbation spectrum can arise 
depending on the
 sign and magnitude of $c$. After inflation $|c|$ 
need not be small compared with 1 (in the case where $c < 0$) 
and the $cH^{2}$ term can even widen the range of 
expansion rate and energy density during inflation which is compatible with the
observed density perturbations. 
\newline {\bf (ii) Non-renormalisable superpotential corrections}

           We next consider adding Planck scale-suppressed 
non-renormalisable terms (NRTs) to the r.h. sneutrino superpotential, 
\be{e40} W_{N} = \frac{1}{2}M_{N} N^{2} + \frac{\lambda_{n} N^{n}}{n! 
M^{n-3}}  ~,\ee
where $n!$ is a symmetry factor and $\lambda_{n} \approx 1$.
For large enough $|N|$ the scalar potential becomes dominated by the NRT contribution, 
\be{e41}  V(N)  \approx  \frac{\lambda_{n}^{2} N^{2\left(n-1\right)}}{2^{n-1} \left(n-
1\right)!^{2} M^{2\left(n-3\right)} }   ~.\ee
The effect of the NRTs is to place an upper limit on the value of $N_{osc}$. 
During inflation, the effective mass $V^{''}(N)$ must be small compared with 
$H^{2}$. After inflation and during IMD, 
$H^{2} \propto a^{-3}$. Therefore $V^{''}(N)$ may become larger than $H^{2}$, at 
which point the $N$ field will begin to evolve.
$N$ will then track the value at which $V^{''}(N) \approx H^{2}$, since 
as $H^{2}$ decreases below $V^{''}(N)$, the rate of roll of $N$ will increase until the 
rate of decrease of $V^{''}(N)$ matches rate of the decrease of $H^{2}$. 
In particular, when $N$ will first begins to slow-roll we have 
$3H\dot{N} \approx -V^{'}(N)$. For $V(N) \propto N^{2\left(n-1\right)}$, this has a 
solution $N \propto a^{3/(4-2n)}$, such that $V^{''}(N) \propto a^{-3} \propto H^{2}$. So 
the slow-rolling solution will be such that $V^{''}(N)$ tracks $H^{2}$ and is of the 
same order of magnitude as $H^{2}$. 
Thus the value of $N$ at which $V^{''}(N) \approx H^{2}$ places an upper 
limit on 
$N$ for a given value of $H$, $N_{lim}$, given by 
\be{e42} \frac{N_{lim}}{M} \approx \frac{\alpha_{n}}{\lambda_{n}^{\frac{1}{n-2}}}  
\left(\frac{H}{M}\right)^{\frac{1}{n-2}}   ~,\ee 
where $\alpha_{n}$ is a constant of order 1. 
Therefore for $N_{lim}$ to be greater than $N_{osc}$ at $H_{osc}$ we require 
that $n$ is greater than $n_{lim}$, where
\be{e43} n_{lim} = 2 + \frac{{\rm ln}\left(\frac{H_{osc}}{M}\right)}{{\rm 
ln}\left(\frac{\lambda_{n}^{\frac{1}{n-2}}N_{osc}}{\alpha_{n} M}\right)}    ~.\ee

      For the case of Dirac neutrino masses, we require from the late decay condition 
that $N_{osc}/M \gae 1 \times 10^{-5}$. Thus with 
$H_{osc} \approx m_{\tilde{N}} = 100 \GeV$, $N_{osc}/M \approx 1 \times 10^{-5}$ 
and $\lambda_{n} \approx \alpha_{n} \approx 1$, we require that
$n > n_{lim} = 5.3$. Thus a suppression of Planck-scale suppressed 
NRTs in the r.h. neutrino superpotential of dimension less than 6  
is required in this case.  This might be achieved by a modest discrete 
symmetry. 

    For the case of Majorana neutrino masses, with 
$H_{osc} \approx m_{\tilde{N}} = 100 \GeV$, $N_{osc}/M \approx 0.01$ 
and $\lambda_{n} \approx \alpha_{n} \approx 1$, we require that
$n > n_{lim} = 10.2$. 
This is a significant problem for a Majorana r.h. sneutrino curvaton. It requires a  
high degree of suppression of Planck-suppressed non-renormalisable terms, eliminating all 
NRTs in the r.h. neutrino superpotential up to $n < 11$. However, as we will discuss, 
there is likely to be a more severe problem for 
the Majorana r.h. sneutrino curvaton, namely the 
survival of the r.h. sneutrino condensate in the inflaton decay product background.

     Although Planck scale-suppressed NRTs appear to disfavour a r.h. sneutrino 
curvaton, it should be noted that conventional SUSY inflation models also have  
problems with Planck scale-suppressed NRTs. Chaotic inflation models require that the 
inflaton expectation value is greater than $M$ \cite{chaotic}, whilst SUSY hybrid 
inflation models require that $N$ is close to 
$M$ when scales corresponding to observed cosmic microwave background (CMB) perturbations exit 
the horizon (assuming natural values of the renormalisable gauge and Yukawa 
couplings) \cite{dti,fti}. 
 In this sense the r.h. sneutrino curvaton may be no more 
problematical that conventional SUSY inflation models with respect to Planck 
scale-suppressed superpotential terms.

\subsection{Decay temperature of the Dirac and Majorana r.h. sneutrino condensate}

The neutrino mass is related to the  Yukawa coupling by 
$m_{\nu} = \lambda_{\nu} v_{u}$. The r.h. sneutrino mass is simply given by the 
SUSY breaking mass term, $m_{\tilde{N}} = m_{o} \approx 100 \GeV$. 
Thus the 
temperature of the Universe when the condensate decays, 
$T_{\tilde{N} d}$, is given by
$\Gamma_{\tilde{N} d} \approx H(T_{\tilde{N} d})$, 
where $H(T) = k_{T}T^{2}/M_{Pl}$ and 
$k_{T} = (4 \pi^{3} g(T)/45)^{1/2}$. Thus  
with the r.h. sneutrino decay rate given by \eq{e19}, the 
temperature of r.h. sneutrino decay is, in general, 
\be{e44x} T_{\tilde{N} d} = \left(\frac{\lambda_{\nu}^{2} m_{\tilde{N}}M_{Pl}}{4 \pi 
k_{T_{\tilde{N} d}}}
\right)^{1/2}   ~.\ee 
For the case of the Dirac r.h. sneutrinos, \eq{e44x} implies that
\be{e44} T_{\tilde{N} d} = \left(\frac{m_{\nu}^{2}m_{\tilde{N}}M_{Pl}}{4 \pi 
v_{u}^{2} k_{T_{\tilde{N} d}}}
\right)^{1/2} = 4.4 
\left(\frac{m_{\nu}}{0.1 \eV}\right)\left(\frac{m_{\tilde{N}}}{100 \GeV}\right)^{1/2} \MeV  
~,\ee
where we have used $v_{u} \approx 100 \GeV$ and $k_{T_{\tilde{N} d}} \approx 5$, corresponding to 
$\gamma$, $e^{\pm}$ and  $\nu_{i}$ ($i = 1,2,3$) as light degrees of freedom \cite{tk}.
Thus the Dirac r.h. sneutrino condensate typically decays in the temperature range $1-10 \MeV$.
(Note that if the energy density of the r.h. sneutrino condensate 
dominates the energy density of the Universe when it decays,  
$T_{\tilde{N} d}$ should then be interpreted as the temperature to which the 
Universe reheats after the condensate decays.)  
Since the Dirac r.h. sneutrino condensate decays well below the temperature 
at which weakly interacting particles of mass of the order of $m_{W}$
freeze out of chemical equilibrium ($T_{freeze} \approx 1-10 \GeV$), 
an important constraint on the Dirac r.h. sneutrino curvaton 
is the requirement 
that the lightest supersymmetric particles (LSPs) 
produced in decay of the r.h. sneutrino condensate do 
not overclose the Universe. This requires that the LSPs decay 
before nucleosynthesis at $T \approx 1 \MeV$, in order that the light 
element abundances are not disrupted by photo-dissociation due to LSP decay 
cascades \cite{sark}. Therefore the LSP lifetime must satisfy
\be{e47}  \tau_{LSP} \lae H^{-1}(T \approx 1 \MeV)  \approx 1 s    ~.\ee
Thus although the LSP must be unstable its lifetime can be much longer than the time 
required to escape particle detectors ($\sim 10^{-8}$s), in which case experimental searches 
for SUSY particles would be unaffected. A wider range of LSP candidates 
would be allowed than in the R-parity conserving case where the LSP properties are  
constrained by the thermal relic cold dark matter density. This might be testable at future 
colliders such as the CERN Large Hadron Collider. 

    For the case of Majorana r.h. sneutrinos, \eq{e44x} implies that 
\be{e44a} T_{\tilde{N} d} = \left(\frac{m_{\nu}m_{\tilde{N}}^{2}M_{Pl}}{4 \pi 
v_{u}^{2} k_{T_{\tilde{N} d}}}
\right)^{1/2}  = 2.2 \times 10^{3} 
\left(\frac{m_{\nu}}{0.1 \eV}\right)^{1/2} 
\left(\frac{m_{\tilde{N}}}{100 \GeV}\right) \GeV  
~.\ee
Thus since $m_{\tilde{N}}$ typically will not be much larger than $1 \TeV$ 
as a result of the late decay 
condition, \eq{e24}, we expect that $T_{\tilde{N} d} \sim 10^{4} \GeV$ for the 
Majorana r.h. sneutrino condensate.

\subsection{Thermalisation of the right-handed sneutrino condensate}

    We have been assuming throughout the preceeding discussion that the 
r.h. sneutrino condensate survives in the enviroment of the 
inflaton decay products. However, it is possible that r.h. sneutrinos 
in the condensate could be inelastically scattered by collisions
 with the MSSM particles in the 
inflaton decay product background. If the relativistic 
particles in the background have energy $E_{d}$, then the scattering cross-section of a  
r.h. sneutrino, at rest in the condensate, from a relativistic particle in 
the inflaton decay product background via 
Higgs(ino) or (s)lepton  exchange is expected to be 
\be{th1} \sigma_{\tilde{N}sc} \approx \frac{\alpha_{\nu}\alpha_{g}}{E_{CM}^{2}}    ~,\ee
where $\alpha_{\nu} = \lambda_{\nu}^{2}/4 \pi$, $\alpha_{g} = g^{2}/4 \pi$,  where $g$ is a 
typical MSSM gauge/Yukawa coupling and $E_{CM} \approx \sqrt{E_{d}m_{\tilde{N}}}$ 
is the centre of mass energy of the process. The scattering rate from the decay product 
background is then
$\Gamma_{\tilde{N}sc} = n \sigma_{\tilde{N}sc}$, where $n$ is the number density of 
particles in the inflaton decay product background. 
The largest scattering rate will occur for the 
largest $n$ and smallest $E_{d}$, which corresponds to thermalised 
decay products ($T \lae T_{th}$). In this case 
$n = g(T)T^{3}/\pi^{2}$ and $E_{d} \approx T$ \cite{tk}. Therefore  
\be{th2}   \Gamma_{\tilde{N}sc} \approx 
\frac{g(T) \alpha_{\nu}\alpha_{g}}{\pi^{2}} \frac{T^{2}}{m_{\tilde{N}}}     ~.\ee    
Then assuming that thermalisation of the inflaton decay products occurs during RD, the condition 
that the condensate is unthermalised is $\Gamma_{\tilde{N}sc} \lae H(T) \equiv k_{T}T^{2}/M_{Pl}$, 
which implies
 \be{th3}    \lambda_{\nu} 
\lae \left( \frac{4 \pi^{3} k_{T}}{\alpha_{g} g(T)} \right)^{1/2}
\left(\frac{m_{\tilde{N}}}{M_{Pl}}\right)^{1/2}   ~.\ee
For the case of Dirac neutrino masses, this condition becomes
\be{th4}  m_{\nu} \lae \left( \frac{4 \pi^{3} k_{T} 
v_{u}^{2}}{\alpha_{g} g(T)} \right)^{1/2}
\left(\frac{m_{\tilde{N}}}{M_{Pl}}\right)^{1/2} \approx 
\frac{1.0}{\alpha_{g}^{1/2}} 
\left(\frac{m_{\tilde{N}}}{100 \GeV}\right)^{1/2} 
\left(\frac{v_{u}}{100 \GeV}\right)
{\rm keV}    ,\ee
where we have used $g(T) \approx 200$ and $k_{T} \approx 20$. 
Therefore, since $m_{\nu} \approx 0.1$eV,
the Dirac r.h. sneutrino condensate will not be thermalised. 

     For the case of Majorana neutrino masses, the non-thermalisation 
condition becomes
\be{th5} m_{\nu} \lae 
\left( \frac{4 \pi^{3} k_{T}}{\alpha_{g} g(T)} \right)
\frac{v_{u}^{2}}{M_{Pl}} \approx \frac{10^{-5}}{\alpha_{g}} 
\left(\frac{v_{u}}{100 \GeV}\right)^{2} {\rm eV}
~.\ee
Thus typically we expect that this will not be satisfied. For example, if 
we consider $\alpha_{g}$ to correspond to the top quark Yukawa (r.h. sneutrinos 
scattering from thermal top quarks via $H_{u}$ exchange) then $\alpha_{g} \approx 0.1$
and so $m_{\nu} \lae 10^{-4}$eV would be required to evade thermalisation. Therefore 
with $m_{\nu} \approx 0.1 $eV the Majorana r.h. sneutrino condensate will be 
thermalised at $T_{th}$, when the inflaton decay product background thermalises. 

   One possible escape from this conclusion is that the inflaton decay product background 
could remain unthermalised until the energy density of the sneutrino condensate 
comes to dominate the energy density of the Universe. Since 
the Majorana r.h. sneutrino condensate decays at a temperature typically 
around $10^{4} \GeV$,  if $T_{th} \lae 10^{4} \GeV$ then the 
Majorana r.h. sneutrino curvaton may remain a possibility. For the case of 
perturbative inflaton decays, the thermalisation temperature of 
the inflaton decay product background 
$T_{th}$, \eq{e7a}, is proportional to $(T_{R}/m_{S})^{3}$.
Therefore a very 
low background thermalisation temperature is a possibility 
if $T_{R} \ll m_{S}$. 

   The above applies to condensate thermalisation during RD. It is 
straightforward to see that if condensate thermalisation does not 
occur during RD then it will not occur earlier during IMD. If we assume that 
the background is thermalised during IMD (which should give the largest rate of
condensate thermalisation), then since during IMD 
$\Gamma_{\tilde{N}sc} \propto T^{2}$ 
and $H \propto T^{4}$, it follows that once $T > T_{R}$ 
the thermalisation condition $\Gamma_{\tilde{N}sc} \gae H$ 
will become more difficult to satisfy. Thus if the r.h. sneutrino 
condensate does not thermalise during RD it will not thermalise at all.

         In discussing the thermalisation rate, we have implicitly assumed that the mass 
of the $H_{u}$ and $L$ fields coupling directly to the r.h. sneutrino condensate is small 
compared with $T$ and $E_{CM} \approx (T m_{\tilde{N}})^{1/2}$. The effective mass is given by 
$\lambda_{\nu} <N>$, where $<N>$ is the amplitude of the coherent oscillations. In general 
$<N>$ is given by 
\be{em1} <N> = \left(\frac{a_{osc}}{a}\right)^{3/2} N_{osc} 
\approx \left(\frac{T}{T_{R}}\right)^{3/2} \left(\frac{H_{R}}{H_{osc}}\right) N_{osc}  ~.\ee 
Thus for the case of Majorana r.h. sneutrinos the effective mass is given by 
\be{em2} \lambda_{\nu} <N> = \frac{k_{T}T^{3/2} T_{R}^{1/2}}{\sqrt{8 \pi}} 
\left(\frac{m_{\nu}}{m_{\tilde{N}}v_{u}^{2}}\right)^{1/2}  
\left(\frac{N_{osc}}{M}\right)   ~,\ee
which implies that 
$$ \lambda_{\nu} <N> = 4 \times 10^{2} 
\left(\frac{m_{\nu}}{0.1 \eV}\right)^{1/2}
\left(\frac{T}{10^{4} \GeV}\right)^{3/2} $$
\be{em3}  \times \left(\frac{T_{R}}{10^{8} \GeV}\right)^{1/2}
\left(\frac{100 \GeV}{m_{\tilde{N}}}\right)^{1/2}
\left(\frac{N_{osc}}{M}\right) \GeV  ~,\ee 
where we have used $k_{T} \approx 20$ and $v_{u} = 100 \GeV$. 
Thus for $N_{osc}/M < 1$, $m_{\tilde{N}} \gae 100 \GeV$ 
and for a decay temperature of $10^{4} \GeV$ 
for the Majorana r.h. sneutrino, we find that $\lambda_{\nu} <N>$ 
is smaller than $T$ and $E_{CM}$ and so may be neglected when 
considering condensate thermalisation.

       For the case of Dirac r.h. sneutrinos the effective mass is given by 
\be{em4} \lambda_{\nu} <N> = \frac{k_{T}T^{3/2} 
T_{R}^{1/2}m_{\nu}}{\sqrt{8 \pi}v_{u}m_{\tilde{N}} } 
\left(\frac{N_{osc}}{M}\right)   ~,\ee
which implies that
$$ \lambda_{\nu} <N> = 3 \times 10^{-15} 
\left(\frac{m_{\nu}}{0.1 \eV}\right)
\left(\frac{T}{10 \MeV}\right)^{3/2}$$
\be{em3} \times \left(\frac{T_{R}}{10^{8} \GeV}\right)^{1/2}
\left(\frac{100 \GeV}{m_{\tilde{N}}}\right)
\left(\frac{N_{osc}}{M}\right) \GeV  ~.\ee 
Thus the effective mass is generally negligible 
in the case of Dirac r.h. sneutrinos.

\section{Conclusions}

    We have considered the possibility that a r.h. sneutrino could play the role of a 
curvaton in the cosmology of the MSSM extended to accomodate neutrino masses. 

In the case of a Dirac 
r.h. sneutrino, the expectation value of the r.h. sneutrino at the 
onset of its coherent oscillations must satisfy $N_{osc}/M \gae 10^{-5}$, in order that the 
energy density of the r.h. sneutrino condensate dominates the Universe when it decays. 
As a result, Planck-scale corrections to the 
r.h. neutrino superpotential must be suppressed, eliminating all superpotential terms 
$\propto N^{n}$ with $n < 6$. 
$cH^{2}$ corrections to the r.h. sneutrino mass squared 
must also be suppressed during inflation ($|c| \lae 0.1$). 
The inflaton sector of the 
Dirac r.h. sneutrino curvaton scenario can have a much smaller 
expansion rate during inflation than 
conventional SUSY inflation models, with $H_{I} \approx 10^{9} \GeV$ possible. 
(Including the effect of a negative $cH^{2}$ term after inflation, 
values of $H_{I}$ as small as $10^{6} \GeV$ or less are possible.)

     In addition, depending on the sign and magnitude of the 
$cH^{2}$ correction,  
it is possible to have a significantly blue ($c>0$) or 
red ($c < 0$) perturbation spectrum. The recent suggestion from WMAP observations \cite{wmap} 
of a blue perturbation spectrum on comoving scales of the order of $500$ Mpc could therefore be 
accomodated within the curvaton scenario.   

        The late decay of the Dirac r.h. sneutrino condensate 
(at $T \approx 1-10 \MeV$) requires that R-parity be broken and that 
the LSP decays before nucleosynthesis, 
corresponding to a lifetime shorter than 1s.  
(Since this would result 
in the loss of LSP cold dark matter, a new 
dark matter candidate would also be needed.)
Thus the 
Dirac curvaton scenario predicts that LSP properties will typically 
be inconsistent with thermal 
relic cold dark matter. 

   In the case of a Majorana r.h. sneutrino curvaton, we find that the 
requirement that the r.h. sneutrino condensate dominates the energy density of the 
Universe before it decays implies that $N_{osc}/M \gae 0.01$. This imposes a 
strong constraint on Planck-scale suppressed contributions to the 
r.h. neutrino superpotential, requiring elimination of all terms $\propto N^{n}$ 
with $n < 11$. 
However, a more severe problem may arise from scattering of the condensate 
sneutrinos by particles in the thermal background, 
since it is likely that the Majorana r.h. sneutrino 
condensate will be thermalised 
as soon as the inflaton decay product background thermalises. Thus if
the inflaton decay products
thermalise earlier than the time of r.h. sneutrino 
condensate domination of the energy density 
then the Majorana curvaton scenario will be ruled out. Only a sufficiently 
low background thermalisation temperature could evade this conclusion. 
Therefore, although not absolutely excluded, the 
Majorana r.h. sneutrino curvaton scenario 
appears to be more difficult to implement 
that the Dirac curvaton scenario.   

      In conclusion, we find that a r.h. sneutrino
can serve as the source of the density perturbations leading to structure 
formation. 
For the favoured case of a Dirac r.h. sneutrino, 
this requires some suppression of its Planck-scale suppressed superpotential
self-interactions together with sufficiently rapid R-parity violation.
If these conditions are met, the 
Dirac r.h. sneutrino would provide us 
with the only curvaton candidate  
which is strongly motivated by particle physics.

{\it Note Added:} After completing this work we became aware of \cite{mormur} and \cite{postma}, 
which also discuss the case of a Majorana r.h. sneutrino curvaton.

\end{document}